\documentclass[aps,twocolumn,nofootinbib,aps,preprintnumbers,amsmath,amssymb,superscriptaddress,nobalancelastpage,floatfix]{revtex4-1}
\usepackage{graphicx}
\usepackage{color}
\usepackage{bm}
\usepackage{amsmath}
\usepackage{xspace}
\usepackage[normalem]{ulem}
\usepackage{units}
\usepackage{dcolumn}
\usepackage{paralist}
\usepackage{natmove}
\usepackage{natbib}
\usepackage[table]{xcolor}
\usepackage{hyperref}

\begin{document}

\title{Heat and charge transport in bulk semiconductors with interstitial defects}
\author{Vitaly~S.~Proshchenko}
\author{Pratik~P.~Dholabhai}
\author{Sanghamitra~Neogi}
\email{sanghamitra.neogi@colorado.edu}
\affiliation{Ann and H.J. Smead Aerospace Engineering Sciences, University of Colorado Boulder, Boulder, Colorado 80309, USA}

\date{\today}

\begin{abstract}

Interstitial defects are inevitably present in doped semiconductors that enable modern-day electronic, optoelectronic or thermoelectric technologies. Understanding of stability of interstitials and their bonding mechanisms in the silicon lattice was accomplished only recently with the advent of first-principles modeling techniques, supported by powerful experimental methods. However, much less attention has been paid to the effect of different naturally occurring interstitials on the thermal and electrical properties of silicon. In this work, we present a systematic study of the variability of heat and charge transport properties of bulk silicon, in the presence of randomly distributed interstitial defects (Si, Ge, C and Li). We find through atomistic lattice dynamics and molecular dynamics modeling studies that, interstitial defects scatter heat-carrying phonons to suppress thermal transport---1.56\% of randomly distributed Ge and Li interstitials reduce the thermal conductivity of silicon by $\sim$ 30 and 34 times, respectively. Using first principles density functional theory and semi-classical Boltzmann transport theory, we compute electronic transport coefficients of bulk Si with 1.56\% Ge, C, Si and Li interstitials, in hexagonal, tetrahedral, split-interstitial and bond-centered sites. We demonstrate that hexagonal-Si and hexagonal-Ge interstitials minimally impact charge transport. To complete the study, we predict the thermoelectric property of an experimentally realizable bulk Si sample that contains Ge interstitials in different symmetry sites. Our research establishes a direct relationship between the variability of structures dictated by fabrication processes and heat and charge transport properties of silicon. The relationship provides guidance to accurately estimate performance of Si-based materials for various technological applications. 
\end{abstract}

\maketitle

\section{introduction}
Applicability of any material to optical, energy transport or topological architectures is largely determined by our ability to design and manipulate defects and doping in the material that either supply or destroy free carriers. Significant advances have been made to understand the physics and the properties of defects in inorganic semiconductors, owing to the strong historical interaction between the theory of defects and doping-enabled semiconductor-based technologies---microelectronics,~\cite{pacchioni2012defects, ribes2005review, park2012review, wilk2001high} photovoltaics,~\cite{baranowski2016review} light-emitting diodes (LEDs)~\cite{scholz2012semipolar}, and, more recently, spintronics~\cite{weil1984review} and quantum devices~\cite{jelezko2006single, rondin2014magnetometry}. 
Design of defect tolerant semiconductors that can retain their properties despite the presence of crystallographic defects, is actively pursued for the next-generation Earth-abundant solar energy conversion technologies.~\cite{wadia2009materials}
In contrast, introduction of defect-induced innovative transport mechanisms in bulk~\cite{snyder2008complex,shi2016recent} and nanostructured materials~\cite{pichanusakorn2010nanostructured, shakouri2011recent, vineis2010nanostructured, li2010high} enabled paradigm-shifting advances in the thermoelectric (TE) energy conversion devices. Innovative defect engineering efforts in future technology-enabling materials heavily rely on a better understanding of the role of intrinsic point defects in carrier transport. 

In the last two decades, Si-Ge based heterostructures have emerged as key materials in numerous electronic, optoelectronic~\cite{thompson200490,koester2006germanium,liu2010ge} and TE devices.~\cite{chen2003recent, dresselhaus2007new, alam2013review, kissinger2014silicon} During the growth and fabrication of these devices, especially ones that require high temperature processes, several undesirable damages~\cite{srour2003review}---vacancies, interstitials, additional substitutional atoms, clusters---are introduced as a byproduct.~\cite{haque2017overview} These defects mostly degrade the device performances.~\cite{dehlinger2004high,jung2004effect} It is of great technological significance to consider the variability of the configurations, dictated by fabrication processes, while predicting the performance of Si-based materials for various applications. Studies of point defects in Si, introduced through radiation damage, started in the late 1940s.~\cite{srour2003review} However, a unified understanding of stability of interstitial defects and their formation mechanisms was accomplished only recently, with the advent of first-principles modeling techniques,~\cite{chadi1992self,needs1999first,lee1998first, leung1999calculations,rinke2009defect,caliste2007germanium} supported by powerful experimental methods, such as electron paramagnetic resonance spectroscopy and deep-level transient spectroscopy.~\cite{aggarwal1965excitation, watkins1976epr, newman1982defects, watkins2000intrinsic, bourgoin1983point} The commonly studied interstitials in the silicon lattice, using first principles modeling methods, are self-interstitials,~\cite{bar1984siliconSelfInterstitialFirst, needs1999first, lee1998first, wang2004first} germanium,~\cite{wang2004first,caliste2007germanium} carbon,~\cite{watkins1977lattice,burnard1993interstitialC, tersoff1990carbon} and lithium \cite{wan2010first, tritsaris2012diffusion, kim2010nature}. Carbon interstitials play an important role in the radiation-damage behavior of silicon.~\cite{watkins1976epr} Li interstitials in silicon significantly influence the stability of the host material, and thus, the applicability of silicon as an anode material in the next generation lithium-ion batteries.~\cite{kim2010nature} These first principles based electronic structure modeling studies primarily investigated the structural changes, stability and bonding mechanism of defects in the silicon lattice. 

Introduction of vacancies and substitutional point defects have been extensively investigated to suppress the thermal conductivity of bulk silicon.~\cite{joshi2008enhanced, wang2008enhanced, bennett2015efficient} Around a vacancy, lattice relaxation creates additional scattering sites for phonons. The reduction of thermal conductivity is thus explained by increased phonon scattering and a reduction in the mean-free-path of phonons, due to the relatively high concentration of vacancies (1-4\%).~\cite{lee2011effects, wang2014atomistic, bennett2015efficient} However, the number of studies to investigate the influence of interstitials on thermal conductivity is fairly limited. Few studies exist that focused on phonon properties employing Green's function techniques with approximated force constant based models.~\cite{brice1965phonon, ohashi1976resonance, talwar1983dynamical, talwar1987theory, bellomonte1966vibrations} Interstitials in the context of thermoelectrics have been widely studied in the form of clathrates and other caged structures.~\cite{takabatake2014phonon} A comprehensive understanding of the effect of different naturally occurring interstitials (Si, Ge, C and Li) on the thermal and electrical properties of silicon, however, doesn't exist. In this work, we present a systematic study of the variability of charge and heat transport properties of bulk silicon in the presence of randomly distributed interstitial defects (Si, Ge, C and Li), using first principles density functional theory (DFT) and atomistic lattice dynamics and molecular dynamics (MD) techniques, respectively. This theoretical investigation furnishes indirect measures to estimate the presence of interstitial defects in a sample. Furthermore, our study establishes a processing-structure-transport (heat and charge) property map that provides guidelines for design, synthesis and processing to develop Si-based materials with predictable, robust and optimal performance.  

As an illustration of the relevance of this work for practical applications, we discuss the 
relationship between the structure dictated by processing and material performance of bulk silicon, with naturally occurring interstitial impurities in the lattice, for thermoelectric applications. Solid state TE generators~\cite{nolas2013thermoelectrics} are expected to play a key role to meet the rapidly increasing power demands of the internet of things, by converting any source of heat into electricity. Defect engineering in TE materials aims to affect the interdependent heat and charge transport properties toward higher material performance. The maximum power-generation efficiency ($\eta$) of a TE material is, $\eta = \left( \frac{T_{\text{hot}} -T_{\text{cold}}}{T_{\text{hot}}}\right)\left[ \frac{\sqrt{1+ZT}-1}{\sqrt{1+ZT}+\left(\frac{T_{\text{cold}}}{T_{\text{hot}}} \right)}\right]$, where the Carnot efficiency is the ratio of the temperature difference between the hot and cold end $(T_{\text{hot}} -T_{\text{cold}})$ to $T_{\text{hot}}$. The dimensionless figure of merit, $ZT$, is the primary parameter that determines the efficiency and is given by $ZT = \frac{\sigma S^2 T}{\kappa_e + \kappa_{ph}} = \frac{\text{PF}\ T}{\kappa_e + \kappa_{ph}} $, where $\sigma$ is the electrical conductivity, $S$ is the thermopower or Seebeck coefficient, $T$ is the temperature, PF $(=\sigma S^2)$ is the power factor, $\kappa_e$ and $\kappa_{ph}$ are the electronic and ionic contributions to the thermal conductivity, respectively.~\cite{nolas2013thermoelectrics} Our MD studies show that interstitial defects scatter heat-carrying phonons to suppress the $\kappa_{ph}$; 1.56\% of randomly distributed Li, Ge, Si and C interstitials reduce the thermal conductivity of silicon by $\sim$ 34, 30, 20 and 9 times, respectively. We provide further insights into the modification of phonon propagation properties in the presence of interstitials by computing the density of states and group velocities using lattice dynamics technique. Defects commonly introduce additional impurity levels within the TE material’s energy band gap,~\cite{grimmeiss1977deep, song2016effect} leading to a reduced charge transport. We investigate the electronic transport properties of bulk Si with 1.56\% Ge, C, Si and Li interstitials in hexagonal, tetrahedral, split-interstitial and bond-centered sites. We demonstrate that Si and Ge interstitials in the hexagonal sites minimally impact the charge transport properties of bulk silicon. This combined with the decreased values of $\kappa_{ph}$ leads to a 14 and 17 times improved $ZT$, respectively. To complete the study, we predict the thermoelectric property of an experimentally realizable bulk silicon sample with Ge interstitials in different symmetry locations.  

\section{Method}
In order to elucidate the impact of interstitial defects on the electronic and heat transport properties of bulk silicon, we employed first principles DFT and atomistic lattice dynamics and MD techniques, respectively.
\subsection{System details}
\begin{figure}
\includegraphics[width=1.0\linewidth]{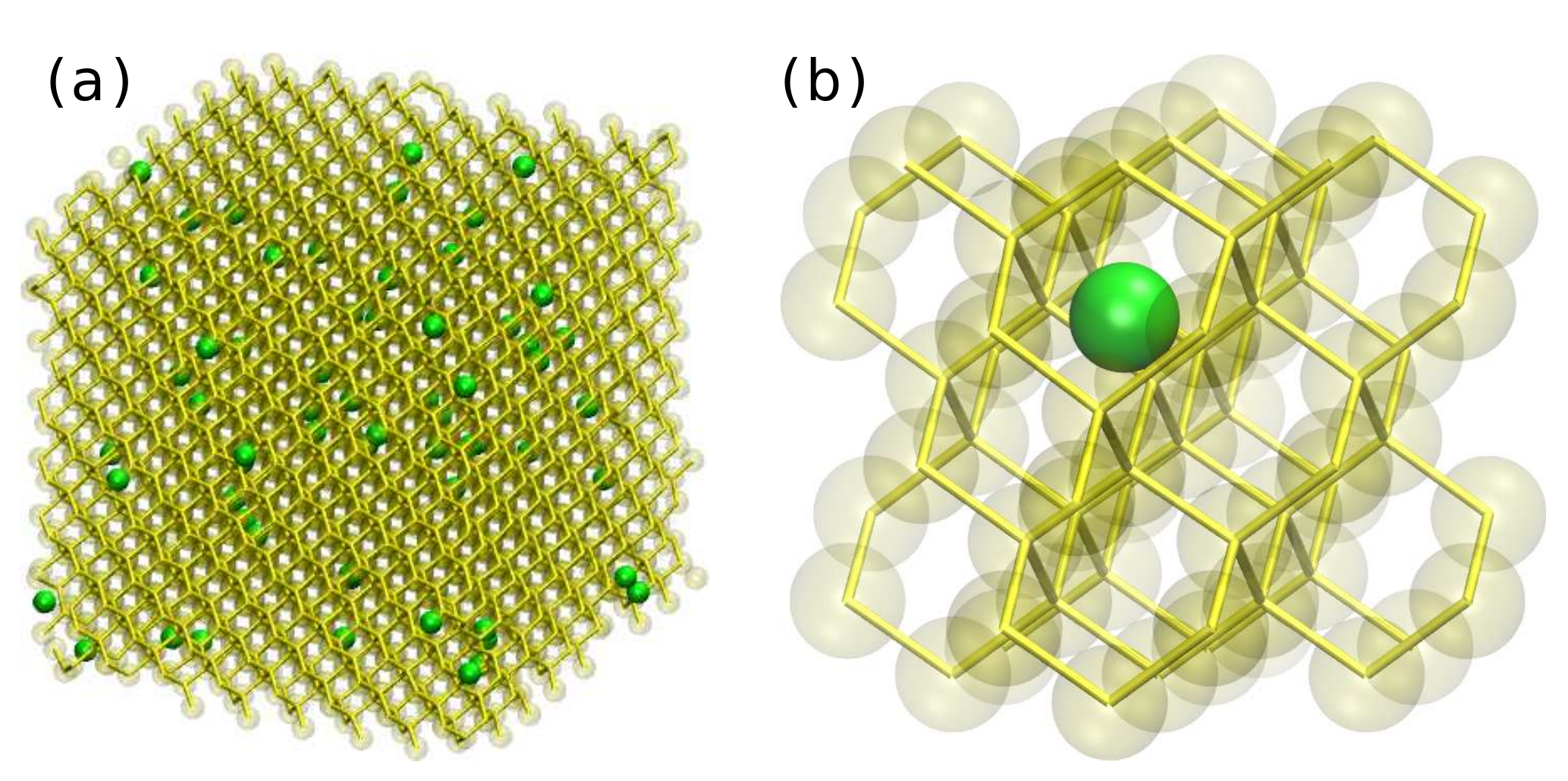}
\caption{Representative configurations of bulk Si with Ge interstitials studied using (a) classical MD and (b) first-principles DFT method. The MD configuration consists of 4096 host Si (yellow) and 64 interstitial Ge (green) atoms. The DFT configuration has 64 Si (yellow) and 1 Ge (green) atoms. The concentration of Ge in bulk Si is 1.56\% in both configurations.}
\label{fig:system}
\end{figure}

\begin{figure}
\includegraphics[width=1.0\linewidth]{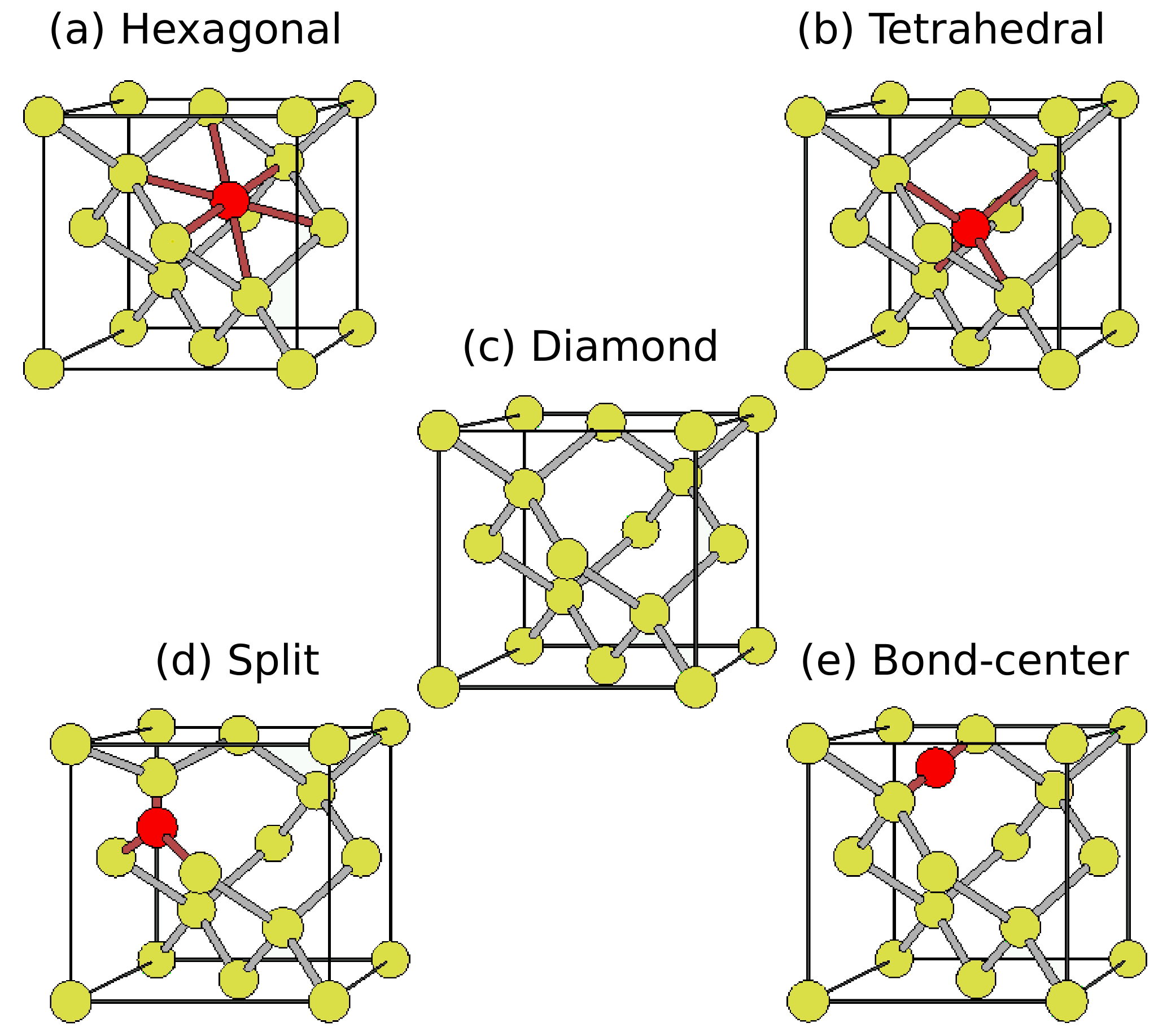}
\caption{(c) Conventional unit cell of bulk silicon and commonly studied interstitial sites (a, b, d and e) in the diamond cubic lattice. Yellow and red colors represent host and interstitial atoms, respectively.}
\label{fig:interstitialsType}
\end{figure}

We investigated bulk silicon configurations with commonly occurring interstitial impurities: Ge, C, Si, and Li; the systems studied will henceforth be referred to as Si-I-X systems (where I $\equiv$ Interstitials and X = Ge, C, Si or Li). Representative microscopic configurations of the Si-I-X systems investigated are shown in Fig.~\ref{fig:system}. 
The bulk Si supercell templates were generated by replicating a conventional unit cell (CC) of silicon, which consists of 8 atoms in a diamond face centered cubic lattice arrangement with side length 5.431 {\AA}. The guest atom species (X) were then inserted in the interstitial sites (I) of these supercell templates to create the Si-I-X configurations. The most widely studied interstitial sites in bulk silicon lattice, investigated with semi-empirical and first-principle methods, are of hexagonal, tetrahedral, split-interstitial, and bond-centered type.~\cite{bar1984siliconSelfInterstitialFirst, needs1999first, lee1998first, wang2004first,caliste2007germanium, watkins1977lattice,burnard1993interstitialC, wan2010first,tritsaris2012diffusion} Previous studies discussed that each of these sites is energetically favorable for only a specific set of guest atoms as interstitials. Additionally, the stability of interstitials also depends on the charge state of the defects due to the change in the number of dangling bonds.~\cite{lee1998first,chadi1992self} We investigated electronic transport properties of relaxed Si-I-X configurations with neutrally charged interstitials in each of these stable sites (illustrated in Fig.~\ref{fig:interstitialsType}) using DFT. 
Each of the $[2 \times 2 \times 2] \times \textrm{CC}$ DFT model supercell configurations (Fig.~\ref{fig:system}) has 64 Si and 1 interstitial (I) atoms yielding a 1.56 \% concentration of guest atoms in host Si. The system size was chosen to allow us to study charge transport in a large set of defected semiconductors with a varied chemical bonding environment. In parallel, to study the extent of disruption of phonon transport in the presence of interstitial scatterers, we performed MD studies of the Si-I-X systems with three different interstitial concentrations, 0.1\%, 0.5\%, and 1.56\%. The MD models consisted of several thousands of atoms ($\geq 32768$). The system sizes were chosen to represent bulk systems with low enough interstitial concentrations realizable in experimental conditions,~\cite{kissinger2014silicon} as well as to perform a large set of calculations with reasonable computing expenses. The interstitial sites in these large replicated MD supercells were generated using a systematic search algorithm. The algorithm employs a repulsive pair potential to model interatomic interactions between an interstitial atom and its neighboring host lattice atoms to identify interstices in materials.~\cite{jiang2008computational} These generated tetrahedral sites were randomly populated with a chosen number of interstitial atoms consistent with the concentration. Three distinct randomized configurations were investigated for each concentration. We obtain the initial configurations of the Si-I-X systems by placing all the different species investigated (Ge, C, Si, and Li) at tetrahedral interstitial sites, to ensure a consistent comparison of their impact on thermal properties of bulk Si. However, as we discuss in the following ``Atomistic calculations" subsection, a percentage of these interstitial atoms are likely to populate other sites in the final configurations following equilibration at 300K, due to thermal fluctuations.

\subsection{Atomistic calculations}
We analyzed the thermal transport properties of Si-I-X systems by performing equilibrium molecular dynamics (EMD) simulations, with periodic boundary conditions enforced in all three directions to emulate infinite systems. The interatomic forces between Si, Ge and C atoms were modeled using the empirical potential proposed by Tersoff.~\cite{tersoff1989modeling} This potential was parametrized to treat heteronuclear bonds, and to reproduce correctly the elastic properties of silicon, SiC and its defects, in particular. To model interactions in the Si-I-Li systems, we implemented a second nearest-neighbor modified embedded atom method (2NN MEAM) interatomic potential, which was used to describe interactions in Li--Si alloys~\cite{cui2012second}. To the best of our knowledge, this potential has not been employed for thermal property calculations. We performed several test calculations of thermal conductivities (TC) of bulk Si with the 2NN MEAM potential, and compared with the TC results obtained with the widely used Tersoff potential. We present the computational details and the comparison, in the following paragraphs. 
 
All systems with interstitials were equilibrated at 300 K to ensure that the stability of the structures is maintained. After equilibration, the initial velocities in the MD simulations were set to $300$K and the systems were coupled to a Nos\'e--Hoover thermostat for 1 ns to decorrelate the systems from their initial configurations. The thermostat was then decoupled from the systems so that the simulations were performed under microcanonical conditions. The equations of motions were integrated with a time step of 0.25 fs to ensure energy conservation over simulation times of several tens of nanoseconds. We computed the TCs from the fluctuations of the heat current in EMD simulations, using the Green--Kubo relation~\cite{zwanzig1965time}: $\kappa_\alpha=1/({\it k}_{B}{\it V}{\it T}^2) \int_{0}^{\infty}{\it dt}\langle{\it J}_{\alpha}({\it t}){\it J}_{\alpha}({0})\rangle$ where $\alpha$ = x, y, z; ${\it k}_{B}$ is the Boltzmann constant, ${\it V}$ is the volume of the system, ${\it T}$ is the temperature, and $\langle{\it J}_{\alpha}({\it t}){\it J}_{\alpha}({0})\rangle$ is the heat current autocorrelation function along the direction ($\alpha$) of heat propagation. The heat current is computed by ${\bf J} = \sum_i^N \epsilon_i {\bf v}_i + 1/2 \sum_{i,j;i\neq j}^N ({\bf F}_{ij}\cdot {\bf v}_i){\bf r}_{ij}+1/6 \sum_{i,j,k;i\neq j;j\neq k}^N ({\bf F}_{ijk}\cdot {\bf v}_i)({\bf r}_{ij}+{\bf r}_{ik})$, where $\epsilon_i$ and ${\bf v}_i$ are the energy density and velocity associated with atom $i$, respectively. ${\bf F}$ is the interatomic force acting between atoms separated by a distance ${\bf r}$. The heat flux data were recorded at 5 fs intervals. The total simulation times varied between 15 and 30 ns for different configurations. All EMD simulations were carried out using LAMMPS software~\cite{plimpton1995fast}. 

\begin{figure}
\includegraphics[width=1.0\linewidth]{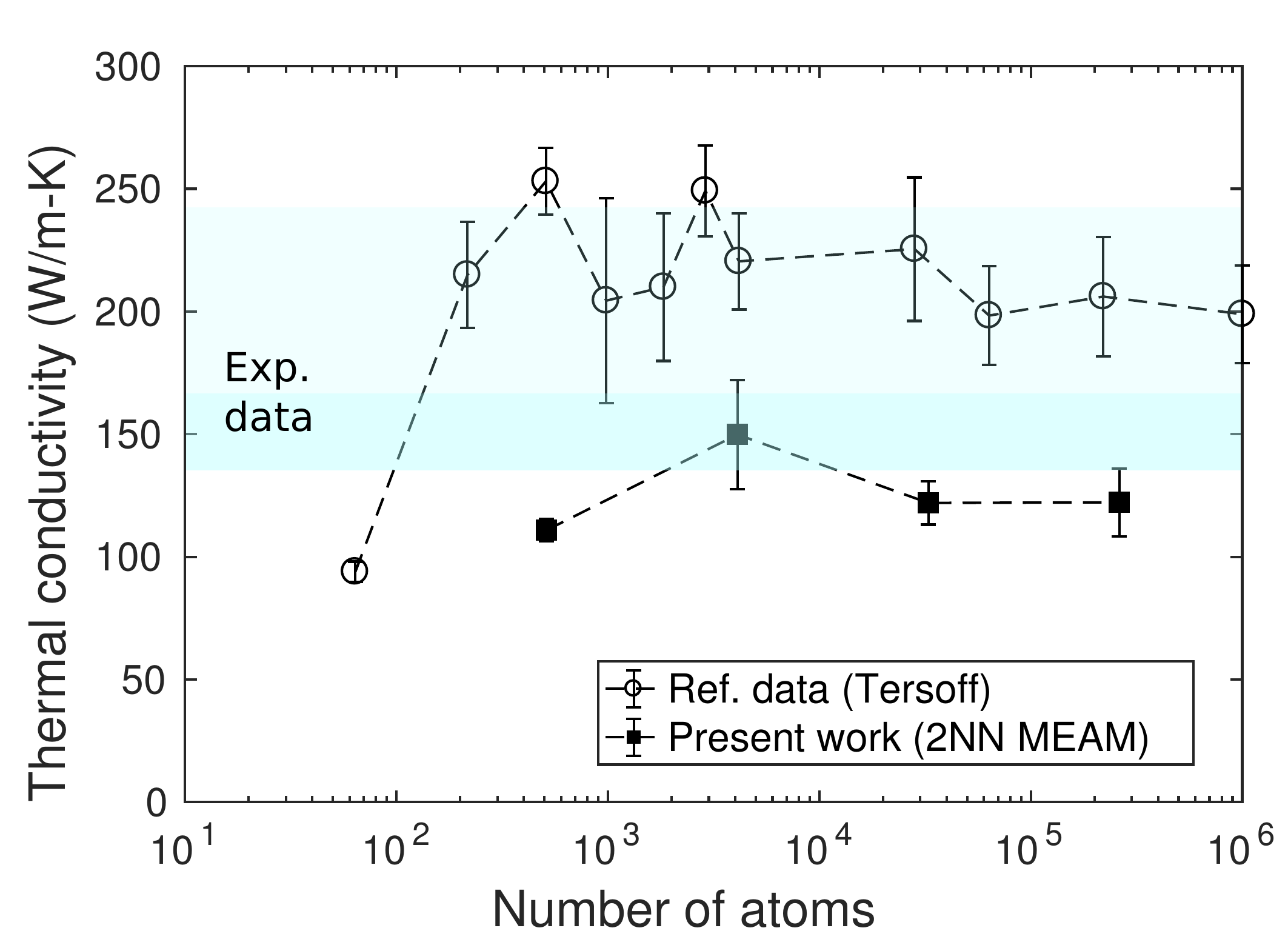}
\caption{Thermal conductivities of bulk silicon at 300 K computed from equilibrium molecular dynamics using the Green--Kubo theorem, as a function of the number of atoms in the simulation cell. The filled squares represent the TCs computed using the 2NN MEAM potential.~\cite{cui2012second} To facilitate a coherent comparison, corresponding TCs obtained with Tersoff potential~\cite{he2012lattice} are shown with open circles (data extracted from Ref.~\cite{he2012lattice}; converged TC: $196.8 \pm 33.3$ W/m-K) alongside 2NN MEAM values. The dashed line connecting the points is displayed as a guide to the eye. The blue shaded region, extracted from Ref.~\cite{he2012lattice}, indicates the range of TCs measured in the experiments where most values lie in the darker portion of the region. The converged TC of bulk Si using 2NN MEAM is $122.22 \pm 13.81$ W/m-K.}
\label{fig:meam}
\end{figure}

A known disadvantage of the Green-Kubo formalism combined with EMD method is that the TCs obtained might suffer from system size dependence. The size converged TC value of bulk Si modeled with Tersoff empirical potential is reported to be $196.8 \pm 33.3$ W/m-K.~\cite{he2012lattice} However, no such study has been reported for Si described with the 2NN MEAM interatomic potential.~\cite{cui2012second} To obtain a reference bulk Si TC described with 2NN MEAM interatomic potential, we performed simulations of supercells of increasing volume from $[4 \times 4 \times 4] \times \textrm{CC}$ to a size $[32 \times 32 \times 32] \times \textrm{CC}$ containing 262144 atoms. As depicted in Fig.~\ref{fig:meam}, we find that the converged bulk Si TC using 2NN MEAM is $122.22 \pm 13.81$ W/m-K. We will use this value in the following analysis as our reference for bulk Si TC, using 2NN MEAM empirical potential. We find that to obtain a converged value of TC, one has to use simulation cells with at least ∼32,768 atoms. Each of the data points reported in Fig.~\ref{fig:meam} was obtained by averaging over 10 calculations performed with independent configurations, and the standard deviation is reported as the uncertainty. Compared to the experimentally measured TC values (depicted in the blue shaded regions in Fig.~\ref{fig:meam}) ranging from 130 to 150 W/m-K at 300 K, it can be argued that the bulk Si TC modeled using 2NN MEAM is in reasonable agreement with the measured values, and the empirical potential offers a rational choice to model bulk Si with Li interstitials.

In order to test system size dependence of TCs of Si-I-X samples, we computed TCs of two systems, Si-I-Ge and Si-I-Li, each with two supercells of sizes $[8\times8\times8]\times\text{CC}$ (4096 atoms) and $[16\times16\times16]\times\text{CC}$ (32768 atoms). This strategy ensured that we tested the size convergence of TCs of defected Si systems described with two different potentials used in this study, Tersoff and 2NN MEAM. For the two system sizes of Si-I-Ge and Si-I-Li, the TCs differ by $\sim$9.8\% and $\sim$9.6\%, respectively, which fall within our method uncertainty limits. In our conservative estimate the dimension of simulation supercell with minimum system size effects is $[16\times 16 \times 16] \times \text{CC}$. Henceforth, all TC computations of Si-I-X systems were performed using this system size. For each of the three chosen concentrations of interstitials (0.1\%, 0.5\%, and 1.56\%) in the Si-I-X systems, TCs are averaged over 45 data sets---obtained from 3 randomized configurations, simulations initialized with 5 different random velocities and results averaged over x, y and z directions. 

\subsection{First principle calculations}

A century of developments in empirical and semi-empirical models~\cite{stoneham2001theory, pantelides1978electronic} realized the modern theory of defects in crystalline solids, based on first-principles electronic structure techniques.~\cite{freysoldt2014first} The method has been demonstrated to expose phenomena that can be readily implemented to optimize the performance of a broad-range of technology-enabling materials. We implemented electronic structure calculations with DFT, using the generalized gradient approximation (GGA) of the exchange-correlation functional by Pedrew-Burke-Ernzerhof (PBE)~\cite{perdew1996generalized} as implemented in the plane-waves code Quantum Espresso (QE),~\cite{giannozzi2009quantum} to study the effect of interstitial species types and their positions on electronic transport. We employed scalar relativistic normconserving pseudopotentials to treat core electrons of both the host Si and the interstitial atoms.~\cite{giannozzi2009quantum} The Kohn-Sham orbitals expanded in terms of a plane wave basis set, had a cutoff energy of 30 Ry for all calculations. A convergence threshold for self-consistency was chosen to be $10^{-9}$. As illustrated in Fig.~\ref{fig:system}(b), the periodic DFT supercells consisted of 64 silicon and 1 interstitial atoms---corresponding to 1.56\% interstitial concentration in the bulk (see the ``System details'' subsection for details). The supercells were first relaxed employing Broyden-Fletcher-Goldfarb-Shanno quasi-newton algorithm with a $4 \times 4 \times 4$ Monkhorst-Pack $k$-point mesh~\cite{monkhorst1976special} to optimize the lattice constants and the atom positions of the Si-I-X supercells. Following relaxation, we performed non self-consistent field (NSCF) calculations to obtain the band energies using a dense $k$-point mesh. Such sampling is necessary to converge the calculation of transport coefficients. 

The electronic transport coefficients were evaluated within the framework of semi-classical Boltzmann transport theory~\cite{ziman1960electrons} as implemented in the BoltzTraP code.~\cite{madsen2006boltztrap} The code employs Fourier expansion to realize an analytical representation of the band energies, computed with DFT. Using such representations of the electronic band structure and the knowledge of the density of states, one can evaluate the Seebeck coefficient ($S$), the electrical conductivity ($\sigma$) and electronic thermal conductivity ($\kappa_e$) and the power factor (PF) ($S^2 \sigma$), in the diffusive regime, by integrating the following expressions over the first Brillouin zone~\cite{ashcroftmermin,madsen2006boltztrap}:       
\begin{align}
\mathcal{L}^{(a)} &= \frac{e^2}{V} \int_{BZ} \frac{d^3k}{4\pi^3} \left[\tau({\bf k}) {\bf \nu}({\bf k}) {\bf \nu} ({\bf k}) (\epsilon_{\bf k}-\mu(T))^a \left(-\frac{\partial f_\mu}{\partial \epsilon_{\bf k}}\right) \right] \label{eq:L_integral}\\
\sigma &= \mathcal{L}^{0}, \label{eq:sigma}\\ 
S &= \frac{1}{eT} \mathcal{L}^{1}/\mathcal{L}^{0} \label{eq:seebeck} \\
\kappa_e &= \frac{1}{e^2 T} \left(\mathcal{L}^{2} - (\mathcal{L}^{1})^2/\mathcal{L}^{0} \right) \label{eq:kappa_e}
\end{align}
where $e$ is the electron charge and $V$ the volume of the system. The integrand in the expression $\mathcal{L}^{(a)}$ consists of the electron relaxation time $\tau(\bf k)$, the electron group velocity ${\bf \nu}({\bf k})$, the $a$-th power of the Kohn-Sham energies $\epsilon_{\bf k}$ with respect to the electronic chemical potential $\mu(T)$ and the derivative of the Fermi-Dirac distribution function ($f_{\mu}$). The transport coefficients obtained using Eq. (\ref{eq:L_integral}-\ref{eq:kappa_e}) are tensors: however since transport is isotropic in the bulk, we report only the trace of $S$, $\sigma$ and $\kappa_e$ divided by three. Our tests showed that NSCF calculations performed with a $15\times15 \times15$ $k$-point mesh were sufficient to converge the transport coefficients of the 65 atom Si-I-X systems under investigation. 

It is computationally very expensive to compute the electron relaxation time $\tau(\bf k)$ from first principles for our systems. We adopt the constant relaxation time approximation (RTA), that assumes that the relaxation time $\tau$ depends only on the carrier concentration and is independent of $\epsilon$ and ${\bf k}$. With this approximation $\tau$ can be factored out of the integrals in Eq. \ref{eq:L_integral}, so that it cancels out in the expression of $S$ (Eq. \ref{eq:seebeck}), while it remains as a prefactor in the $\sigma$ and $\kappa_e$ expressions (Eq. \ref{eq:sigma}, \ref{eq:kappa_e}). In RTA $\tau$ can be estimated using the experimental values of electron (hole) mobility~\cite{jacoboni1977review} and effective masses~\cite{goodnick2006computational}, which is known for bulk silicon. We employed a further assumption that small ($\sim$1\%) concentration of interstitials does not affect the electron relaxation times significantly and compute $\sigma$ and $\kappa_e$ using $\tau$ values obtained from bulk silicon data.~\cite{jacoboni1977review,goodnick2006computational} 

\section{results and discussion}

We now turn to discuss the main focus of this work which is to determine how the presence of interstitials affect heat and charge transport properties of bulk silicon. We performed a series of EMD simulations (see the ``Method" section for details) and computed the TCs of Si-I-X systems with varied concentrations of interstitials. After establishing a series of results with MD simulations, we analyzed the bonding environments of the interstitials to analyze and interpret our data. Additionally, we used lattice dynamics (LD) calculations to understand the propagating character of phonons. Lattice dynamics calculations are essential to gain insights into the modification of the phonon properties of bulk silicon with interstitials. In particular, we computed the vibrational density of states of Si with interstitial defects in tetrahedral sites, as well as the propagating character of the phonons of the defected systems. In parallel, we systematically investigated charge transport properties of systems with a wide range of configurations, incorporating various guest species and a number of interstitials symmetry sites for each type. Finally, we combined the phonon and electron transport properties to establish a structure-processing-figure-of-merit map of bulk silicon with interstitials. 

\subsection{Thermal conductivity}
\begin{figure}
\includegraphics[width=1.0\linewidth]{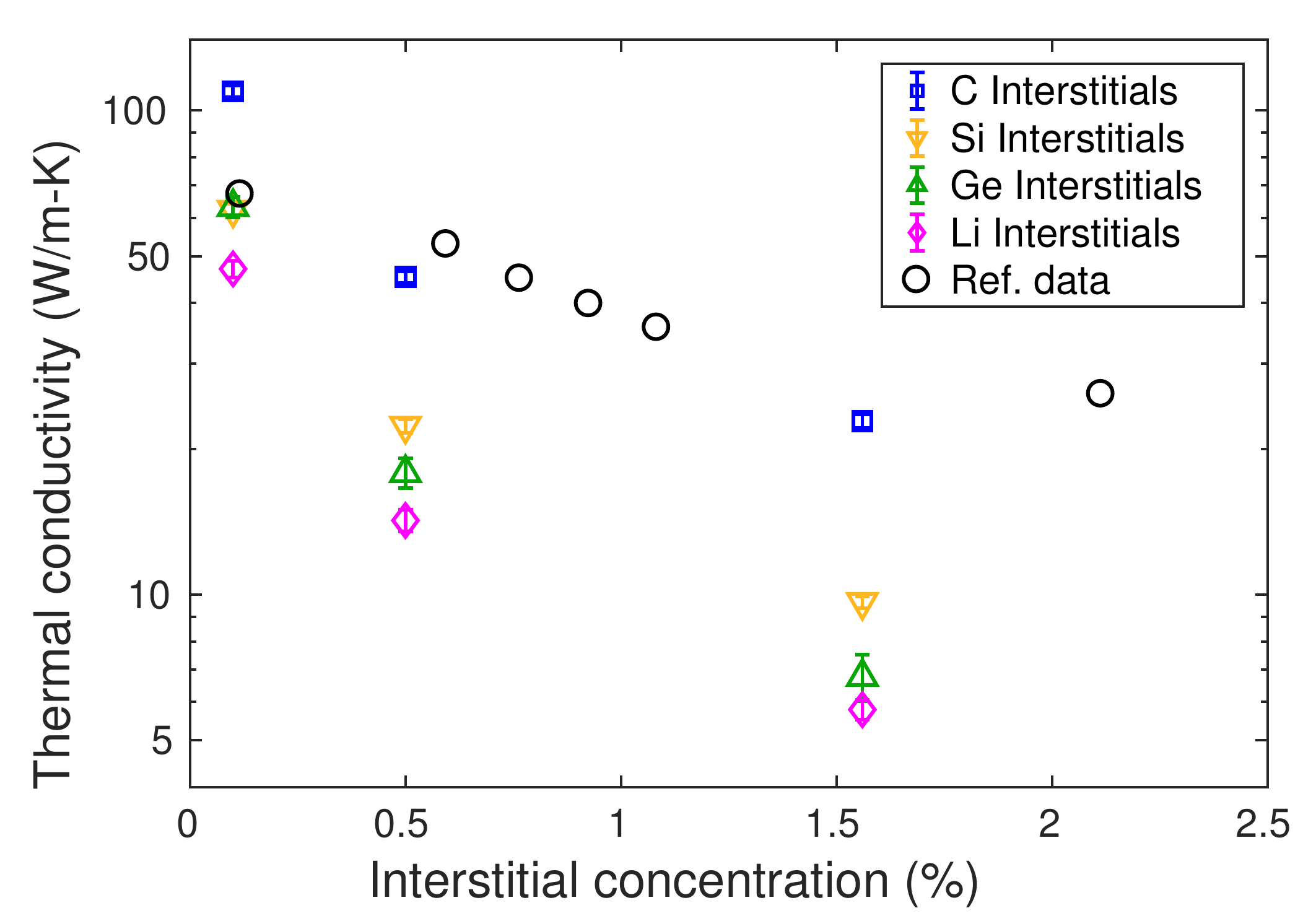}
\caption{Thermal conductivities of bulk silicon with C (blue squares), Si (orange inverted triangles), Ge (green triangles) and Li (magenta diamonds) interstitials at 300 K as a function of interstitial concentrations. To offer a comparison with the effect of widely employed Ge substitutions on TC of bulk Si, corresponding TC values are displayed with black open circles(data extracted from Ref.~\cite{garg2011role}). The X-axis label corresponds to substitutional concentration of Ge in bulk Si in this case.}
\label{fig:kappa}
\end{figure}

We start by discussing our MD results. Figure \ref{fig:kappa} displays the room temperature TCs of Si-I-C (blue squares), -Si (orange inverted triangles), -Ge (green triangles), and -Li (magenta diamonds) systems as a function of interstitial concentrations. The TCs of Si-I-Li systems were computed using the 2NN MEAM potential~\cite{cui2012second} while the TCs of all the other Si-I-X systems were computed with Tersoff potential.~\cite{tersoff1989modeling} However, to facilitate a coherent comparison, we scale the reported values in the plot as follows: TC$^{\text{Li}}_{\text{reported}}$=(TC$^{\text{Li}}_{\text{2NN-MEAM}}/$TC$^{\text{Bulk Si}}_{\text{2NN-MEAM}})\times$TC$^{\text{Bulk Si}}_{\text{Tersoff}}$ where TC$^{\text{Bulk Si}}_{\text{Tersoff}} = 196.8 \pm 33.3$ W/m-K~\cite{he2012lattice} and TC$^{\text{Bulk Si}}_{\text{2NN-MEAM}} = 122.22 \pm 13.81$ W/m-K, respectively. As evident from Fig.~\ref{fig:kappa}, the TC is lowered significantly upon the addition of a small fraction of interstitials, independent of the species type. Remarkably, even with a concentration of 0.1\%, there is a $\sim$ 2 (C)--4 (Li)-fold reduction in the TC from the bulk value ($196.8 \pm 33.3$ W/m-K). The decrease in TC is more pronounced with increased interstitials concentrations, as expected. A general trend of decrease in the TC is noted as Si-I-Li $<$ -Ge $<$ -Si $<$ -C, as displayed in Fig.~\ref{fig:kappa}, with Ge and Li having the most and C the least impact in lowering the TC, respectively. There is almost a factor 2 difference in the TC between systems with different guest atoms with a small ($\sim 0.1\%$) concentration of interstitials. The spread in TC decreases with increased concentration of interstitials---randomly dispersed 1.56\% Li, Ge, Si and C interstitials decrease the TC of bulk Si by $\sim 34, 30, 20$ and $9$ times, respectively. It is interesting to note that the phonon transport in bulk Si is affected more in the presence of Ge interstitials (Fig.~\ref{fig:kappa}) than the widely employed Ge substitutions~\cite{garg2011role} for similar guest species concentrations, by a significant amount. This suggests that interstitials offer a strategy to effectively scatter phonons by tuning the mass and size of the guest atom, and thereby achieve phonon-glass paradigm to engineer next-generation thermoelectric materials.

\subsection{Bonding environment analysis}

\begin{figure}
\includegraphics[width=1.0\linewidth]{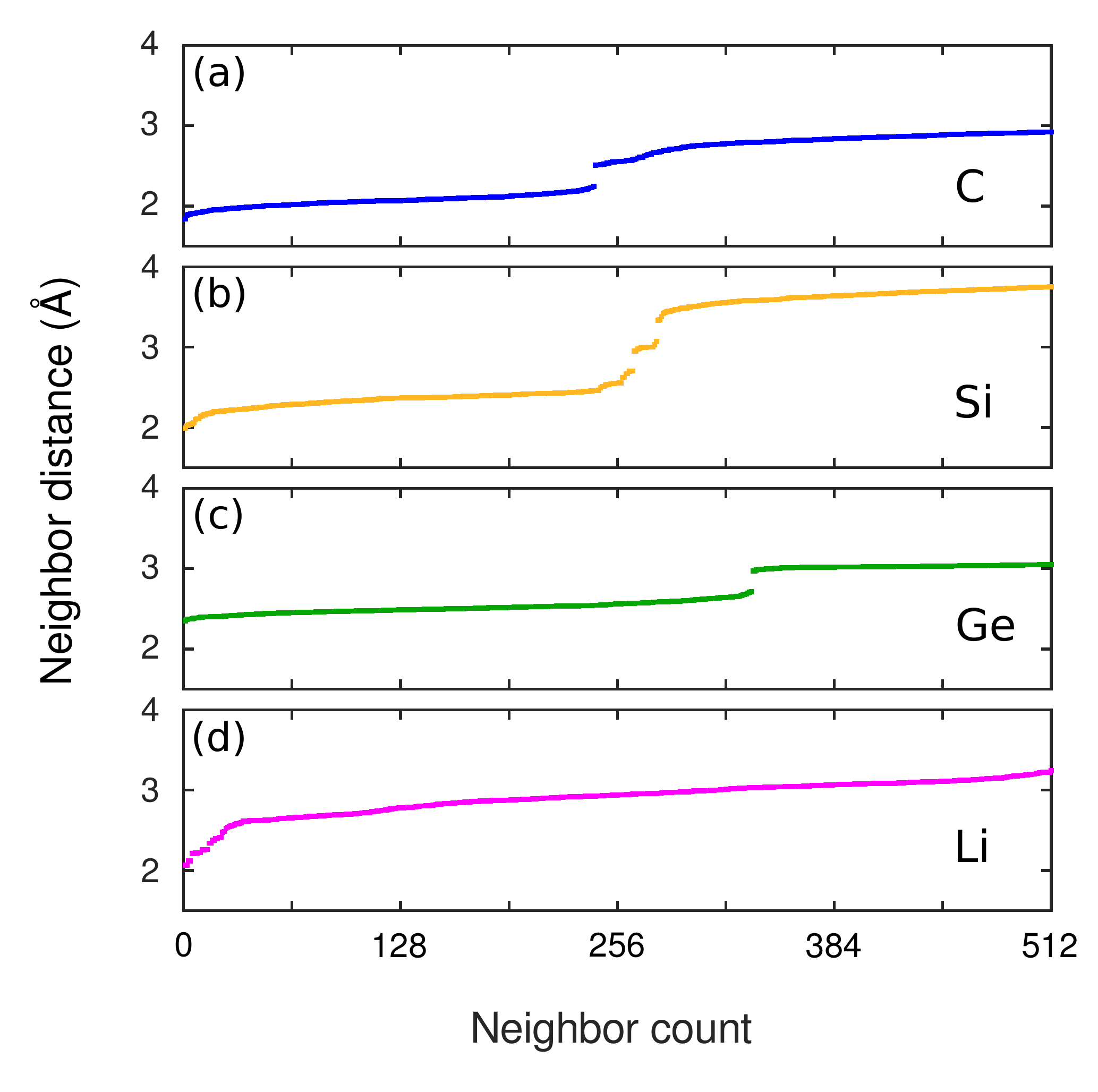}
\caption{Nearest-neighbor environment analysis of Si-I (a) -C, (b) -Si, (c) -Ge, and (d) -Li systems. The neighbor distances are sorted in ascending order over all interstitial atoms in the supercell and displayed against the neighbor count.}
\label{fig:Si-nn}
\end{figure}

To understand the influence of interstitials on phonon transport, we analyzed the bonding environment of the guest atoms in the Si-I-X systems. In Fig.~\ref{fig:Si-nn} we present nearest-neighbor (NN) environment analysis of Si-I (a) -C, (b) -Si, (c) -Ge, and (d) -Li systems. In this analysis, we investigated $[8 \times 8 \times 8] \times \textrm{CC}$ (4096 atoms) Si-I-X supercells with 1.56\% interstitials. This concentration corresponds to 64 guest interstitials in a supercell. For each guest type, we computed the distance between a given interstitial atom and the eight NN Si atoms, to make sure that we include all first NNs in the analysis. A hexagonal interstitial site has maximum number of NNs, 6, in the silicon lattice. Therefore, the total neighbor count is 512 (= $64\times 8$) in the supercell containing 4096 atoms, in a conservative estimate. The neighbor distances are sorted in ascending order over all interstitial atoms in the supercell and displayed in Fig.~\ref{fig:Si-nn}. From the panels (a), (b) and (c) we mark a gap in the neighbor distance curve and infer that this gap indicates the division between the first and the second NNs. The number of the first NNs increases with increasing the size of the interstitial atom (Fig.~\ref{fig:Si-nn})---average first NN distances in Si-I-C, -Si, and -Ge are estimated as  $2.04\pm 0.2$ {\AA}, $2.27\pm 0.28$ {\AA}, $2.53\pm 0.18$ {\AA}, respectively.  

The different number of the first NNs and the uncertainty in the neighbor distances indicate that the interstitials reside in a dynamic bonding environment due to temperature fluctuations. The NN analysis manifests that the interstitials access other symmetry positions during the MD simulations even though the initial configuration consisted of only four-coordinated tetrahedral sites. This was the reason to check 8 NN distances in the previous bonding environment analysis as well. An interesting aspect that can be noted from Fig.~\ref{fig:Si-nn} is that there is no clear separation between the first and the second NN distances in Si-I-Li unlike the other systems investigated. Therefore, we do not report an estimated first NN value for this system. This disparity illustrates that Li interstitials are more diffusive in character leading to significant disorder in the lattice; which is likely to be the reason for Li interstitials being the most effective in lowering bulk Si TC as compared to Ge, C, or Si. One other likely origin of dissimilar NN environment of interstitial atoms is strain. In these materials, atoms with dissimilar masses and sizes compared to the host atoms (other than Si-I-Si) are randomly placed in interstitial sites, leading to anisotropic strain in the material. The strain could further contribute to disparate relaxation that results in diverse bonding environment of the interstitial atoms, leading to varied impact on TC.
\begin{table}[h]
\begin{center}
    \begin{tabular}{ | c | c | c | }
    \hline
    System & Uniform (W/m-K) & Random (W/m-K) \\ \hline
    Si-I-Ge & 11.23 $\pm$ 0.48 & 5.88 $\pm$ 0.23 \\ \hline
    Si-I-Li & 5.62 $\pm$ 0.88 & 5.36 $\pm$ 0.83 \\ \hline
\end{tabular}
\caption[Table caption text]{Thermal conductivities of 4096-atom Si supercells with 1.56\% Ge and Li interstitial atoms placed uniformly and randomly.}
\label{table:distributionTC}
\end{center}
\end{table}

Interstitial atoms potentially serve as scattering centers that reduce the mean free path of phonons. It can be argued that the individual atoms as well as the diverse local structural environment caused by distribution of interstitials, play a critical role in lowering the bulk Si TC. In order to investigate the effect of distribution of interstitials on TC, we investigated two sets of $[8 \times 8 \times 8] \times \textrm{CC}$ (4096 atoms) Si-I-X supercells with 1.56\% Ge and Li interstitials: one set contains configurations in which the interstitials were uniformly placed in tetrahedral sites, the other contains configurations with randomly populated sites. As shown in Table \ref{table:distributionTC}, we find that the TCs of Si-I-Ge-uniform samples are $\sim$2 times higher than the Si-I-Ge-random samples. On the contrary, there is no noticeable difference between the TCs of Si-I-Li-uniform and Si-I-Li-random samples. Our NN analysis shows that the individual bonding environments for the 64 interstitials in the equilibrated (at 300K) Si-I-Ge-uniform samples are similar within statistical fluctuations. However, that is not the case in Si-I-Li-uniform samples due to the diffusive character of Li in Si; the distribution of interstitials in the equilibrated configurations is no longer uniform. Li atoms diffuse through both the uniform and random sample during equilibration to yield a randomized distribution of local strain in the bulk yielding similar TC. Nevertheless, it is important to note that addition of Ge interstitials considerably lowers TC with reduction comparable to Li, despite lacking the diffusive character.  Overall, considering the stability of the material we infer that Ge interstitials offer a promising route to decreasing TC of bulk Si. 

\subsection{Phonon properties}

In order to characterize the transformation of Si heat carriers in the presence of interstitials in Si-I-X systems, and to analyze the TC decrease trend (Ge $<$ Si $<$ C), we carried out a series of lattice dynamics (LD) calculations. In a bulk material, TC can be modeled using the kinetic theory of thermal diffusion, which relies on an approximated solution of the linearized Boltzmann transport equation. The $3\times3$ thermal conductivity tensor can be expressed as~\cite{ziman1960electrons} 
\begin{align}
\kappa_{\alpha \beta} &= \frac{1}{V}\sum_{{\bf q}, \lambda} \hbar \omega_\lambda({\bf q}) \frac{\partial n_\lambda ({\bf q})}{\partial T} v_{\lambda, \alpha}({\bf q}) v_{\lambda, \beta}({\bf q}) \tau_{\lambda}({\bf q}) \label{eq:kappa_ph}
\end{align}
Similar to the expression in Eq. \ref{eq:L_integral} the summation is over all the phonon modes in the first Brillouin zone enumerated by wave vector ${\bf q}$ and polarization $\lambda$. $V$ is the sample volume, $\hbar$ is the reduced Planck constant, $\omega_\lambda({\bf q})$ is the phonon frequency, $n_\lambda ({\bf q})$ is the Bose-Einstein distribution, $T$ is temperature, $v_{\lambda, \alpha}({\bf q})$ is the component of the phonon group velocity vector along the Cartesian direction $\alpha$, and $\tau_{\lambda}({\bf q})$ is the phonon lifetime. Previous studies computed eigenfrequencies of localized phonon modes and phonon density of states of Si with interstitials defects (Si, O, C, B, Li) in tetrahedral sites, using Green's function techniques.~\cite{talwar1983dynamical} However, only isolated interstitials were considered, therefore, the results cannot be compared. We computed the phonon dispersions $\omega_\lambda({\bf q})$ of Si-I-X systems by direct diagonalization of the dynamical matrix of a $8 \times 8 \times 8 \times \textrm{CC}$ (4096 atoms) supercell with 1.56\% randomly dispersed interstitials (as described in the ``System details" subsection). 
\begin{figure}
\includegraphics[width=\linewidth]{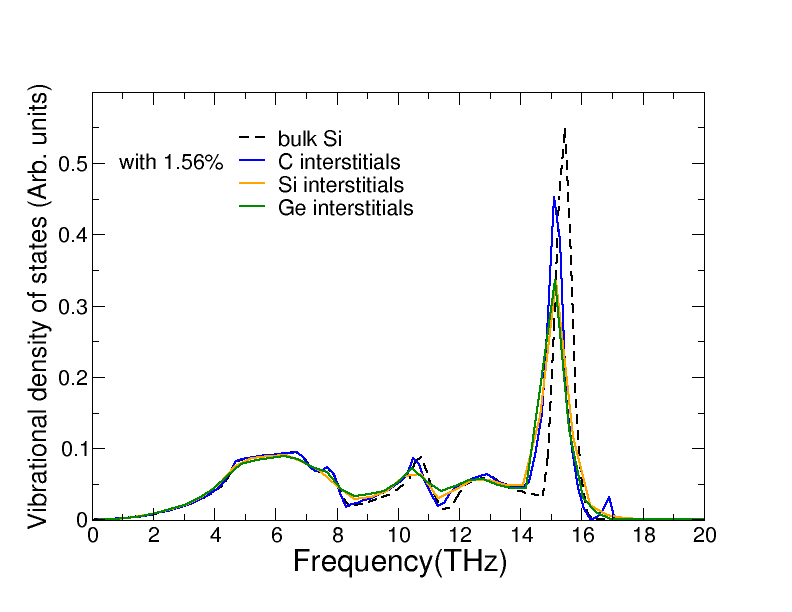}
\caption{Vibrational density of states of bulk silicon with 1.56\% C (blue), Si (orange) and Ge (green) interstitials. 
The population of the phonon modes in Si-I-X systems is slightly affected in the $\sim15$ THz region.}
\label{fig:phonon_dos}
\end{figure}
The sum in equation Eq.~\ref{eq:kappa_ph} is implicitly affected by the vibrational density of states (VDOS) of the system; we computed the VDOS of Si-I-X systems by integrating the first Brillouin zone using a $8\times 8\times 8$ Monkhorst-Pack mesh of $q$-points~\cite{monkhorst1976special}. We find that there is a slight decrease and shift in the intensity of the peak related to optical modes at about 15 THz, as displayed in Fig.~\ref{fig:phonon_dos}, however, these modes do not contribute greatly to heat transport. Previous calculation of phonon density of states also reported a shift in the intensity in the high frequency region, as well as a peak near $\sim 4$ THz. However, only one interstitial defect was considered in this calculation, and simplifying assumptions were made that the interstitials are weakly bound to the host atom, and therefore, should not disturb the local symmetry appreciably.~\cite{brice1965phonon}

In order to analyze the propagating character of the phonons, we compute the group velocities. The phonon group velocity term plays a dominant role in the TC expression (Eq.~\ref{eq:kappa_ph}) and hence, can help us to make a qualitative prediction of TC of the material. The gradient of the frequency $\omega_\lambda({\bf q})$ with respect to wavevector ${\bf q}$ yields the group velocity vector $v_{\lambda}({\bf q})$, i.e., $ v_g = v_{\lambda}({\bf q}) = \partial \omega_\lambda({\bf q})/\partial {\bf q}$.  
\begin{figure}
\includegraphics[width=1.0\linewidth]{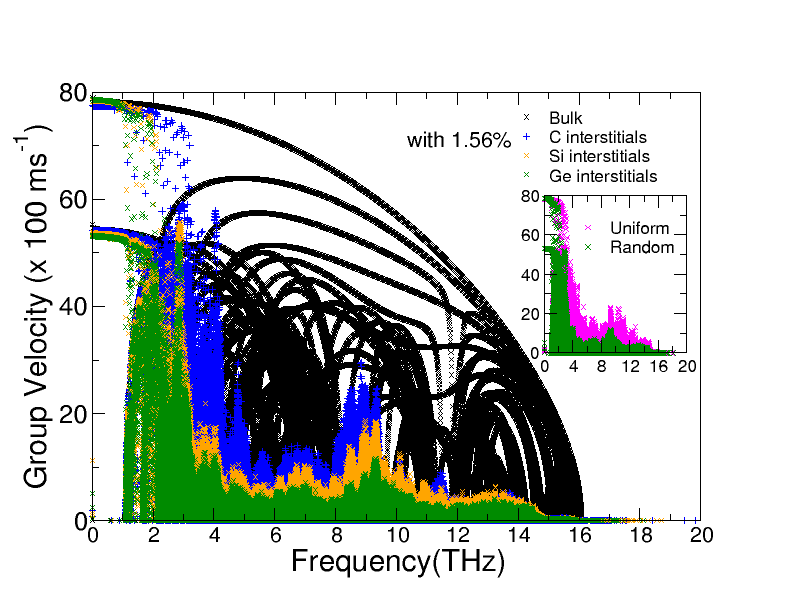}
\caption{Phonon group velocities of bulk silicon with 1.56\% C (blue), Si (orange) and Ge (green) interstitials compared to bulk (black). Phonon group velocities are strongly reduced in the Si-I-X systems above 3--5 THz and the $v_g$ profile is 
approximately of the order Ge $<$ Si $<$ C. Inset: Phonon group velocities in Si-I-Ge systems with uniform (magenta) and random (green) distribution of interstitials. Group velocities are more suppressed in Si-I-Ge system with randomly distributed interstitials.}
\label{fig:phonon_gvel}
\end{figure}
In Fig.~\ref{fig:phonon_gvel}, we show the phonon group velocities in Si-I-C (blue), -Si (orange) and -Ge (green) systems, along the symmetry direction $[000]-[100]$ of the supercell in comparison with bulk (black). We find that the phonon group velocities above 3--5 THz are greatly decreased in defected systems with respect to those in bulk crystalline Si (approximately 1 order of magnitude) due to the presence of interstitials. In fact, the suppression of group velocities has a direct correspondence with the TC decrease trend (Ge $<$ Si $<$ C), as illustrated by our MD simulations (see Fig.~\ref{fig:kappa}). As can be noted from Fig.~\ref{fig:phonon_gvel}, the $v_g$ values in between 3-5 THz are least suppressed in Si-I-C systems, when compared with Si-I-Ge systems. The reduction of $v_g$, especially in the low frequency region, plays a dominant role in heat transport and is directly responsible for the difference in TC values between the systems (Ge $<$ C, see Fig.~\ref{fig:kappa}). The group velocities of Si-I-Si and Si-I-Ge have a similar profile yielding similar TC values. The inset of Fig.~\ref{fig:phonon_gvel} shows the phonon group velocities along the symmetry direction $[000]-[100]$ in Si-I-Ge systems with uniform (magenta) and random (green) distribution of interstitials. The group velocity profile of the Si-I-Ge-random system is lower than the corresponding system with uniform distribution of Ge interstitials, yielding a $\sim$2-fold reduced TC, as reported in Table~\ref{table:distributionTC}. 

\subsection{Electronic transport}
In this section, we illustrate the effect of different interstitials in varied lattice symmetry positions on electronic transport properties of Si. The drastic reduction of TC of bulk silicon with the introduction of interstitials makes this a promising materials engineering strategy for thermoelectric applications. However, in order to achieve a high figure of merit ($ZT$) and attain the phonon-glass-electron-crystal regime, it is important to not only decrease TC but to simultaneously maintain a high power factor (PF). Previous studies with semi-empirical and first-principles methods demonstrated that the energetically stable interstitial sites in bulk silicon are of hexagonal, tetrahedral, split-interstitial and bond-centered type (see Fig.~\ref{fig:interstitialsType}).~\cite{bar1984siliconSelfInterstitialFirst, needs1999first, lee1998first, wang2004first,caliste2007germanium, watkins1977lattice,burnard1993interstitialC, wan2010first,tritsaris2012diffusion} We computed the charge transport properties of the Si-I-X systems (X = Ge, C, Si or Li) with interstitials in each of these possible energetically stable sites. Table \ref{table:energyTable} displays the total energies of the Si-I-X systems with specific atom species X (listed in the leftmost column) in varied interstitial sites of Si, computed with DFT. 
\begin{table}[h]
\begin{center}
    \begin{tabular}{ | c | c | c | c | c |}
    \hline
     & \multicolumn{4}{c|}{Total energies of Si-I-X systems (eV)}\\ \cline{2-5}
    Interstitial & \multicolumn{4}{c|}{Interstitial site}\\
     \cline{2-5}
     atom (X)& Bond-center & Hexagonal & Split  & Tetrahedral   \\ \hline
    Si\cite{bar1984siliconSelfInterstitialFirst, needs1999first, lee1998first, wang2004first} &  \cellcolor{black!30}{}& \cellcolor{blue!10}{0.07} & \cellcolor{blue!10}{0.00} & 0.20   \\ \hline
    Ge\cite{wang2004first,caliste2007germanium} &  \cellcolor{black!30}{}& \cellcolor{blue!10}{0.29} & \cellcolor{blue!10}{0.00} & 0.25   \\ \hline
    C\cite{watkins1977lattice,burnard1993interstitialC} & \cellcolor{blue!10}{0.93} & \cellcolor{blue!10}{0.44} &  0.00 &   \cellcolor{black!30}{}  \\ \hline
    Li\cite{wan2010first,tritsaris2012diffusion} & \cellcolor{black!30}{} & 0.51 &  \cellcolor{black!30}{} & 0.00  \\ \hline
\end{tabular}
\caption[Table caption text]{Total energies of 65-atom Si-I-X supercells with different interstitials in varied sites. The lowest energy across each row is assigned to be the reference and the energies in other columns along the same row is presented with respect to the reference. The gray cells represent systems which are either unstable or have total energies higher than $1$ eV from reference. The blue and white cells correspond to systems with semiconductor-like and metal-like behavior, respectively.}
\label{table:energyTable}
\end{center}
\end{table}
We assigned a reference value (0.00 eV) to the lowest energy across each row and presented the energies in the neighboring columns in the same row with respect to this reference. The split-interstitial sites in Si-I-Si (and -Ge) and Si-I-C systems are considered to be along the [110] and [100] cubic lattice symmetry directions, respectively. The results displayed in Table \ref{table:energyTable} indicate that split-interstitials are the most energetically favorable sites in Si-I-Si, -Ge and -C systems, while Li atoms are more stable in tetrahedral sites at $0$K. These results are in good agreement with previous theoretical calculations. For Si self-interstitials, the split and the hexagonal sites are reported to be the first and the second most stable geometries, respectively, with insignificant difference in energies between them.~\cite{jones2009self, needs1999first, lee1998first, wang2004first} Split-interstitial sites were also demonstrated to be the equilibrium positions for Ge and C defects in the silicon lattice. \cite{wang2004first,caliste2007germanium,watkins1977lattice,burnard1993interstitialC} And, a binding energy analyis reported that Li is most stable in a tetrahedral site.~\cite{wan2010first} However, our MD studies reveal that interstitials are highly likely to access different sites due to temperature fluctuations. Therefore, it is safe to assume that all the sites with energy values reported in Table \ref{table:energyTable} are equally probable due to temperature fluctuations and random defect distributions in a sample. We decided to compute charge transport properties of only the systems that are stable and have total energies within $1$ eV compared to the most energetically favorable system in the same row. The systems represented by the empty gray cells in Table~\ref{table:energyTable} didn't satisfy these criteria and therefore, were omitted from further investigation. Some of the stable Si-I-X systems exhibited semiconductor-like behavior (blue) while other exhibited metal-like behavior (white). We discuss the properties of the semiconductor-like systems in the next paragraphs followed by the metal-like systems. 

\begin{figure}
\includegraphics[width=1.0\linewidth]{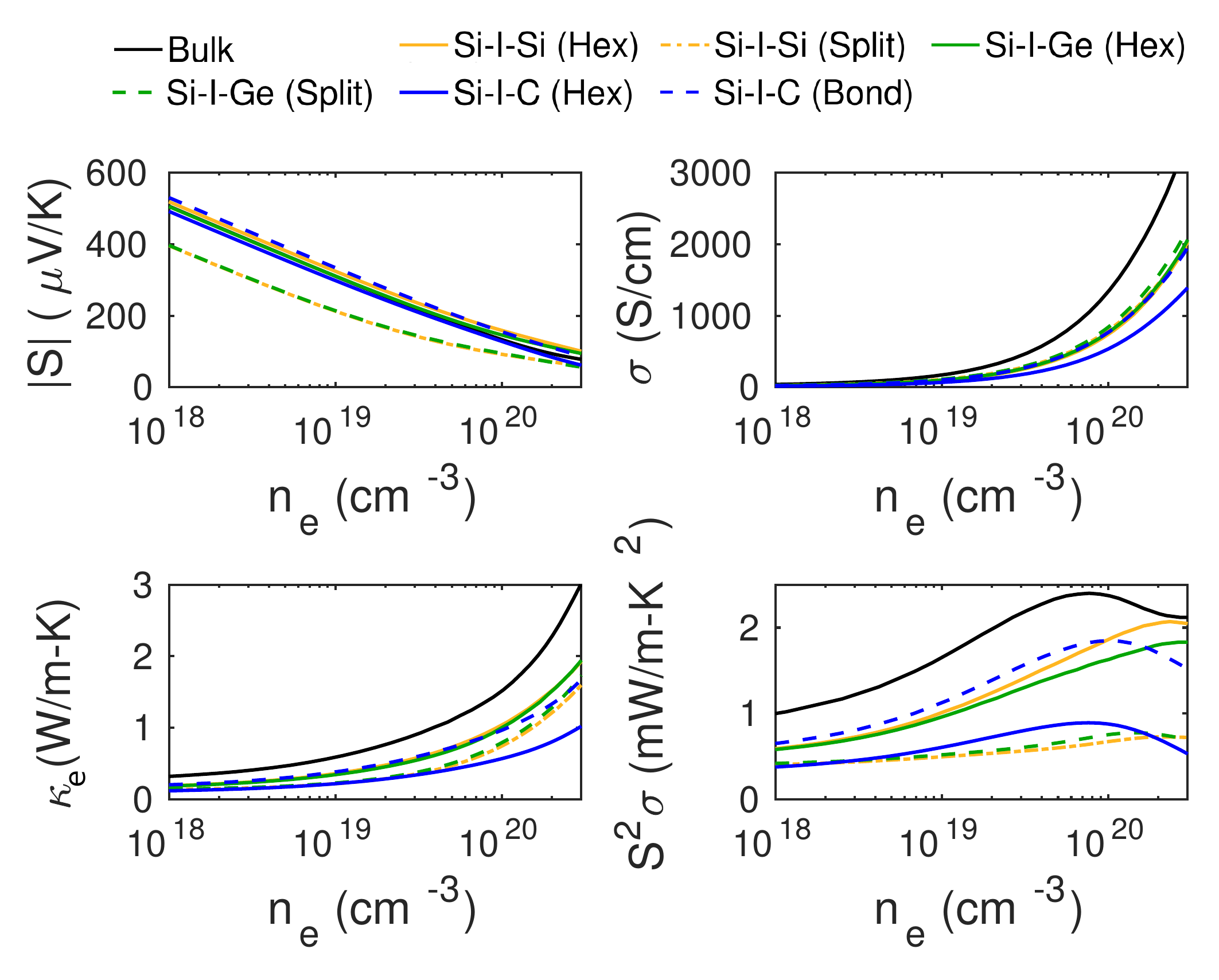}
\caption{Calculated Seebeck coefficient, electrical conductivity, electronic thermal
conductivity and power factor of bulk Si and bulk Si-I-X systems.}
\label{fig:seebeck}
\end{figure}

To illustrate the role of the different interstitials on charge transport, we calculate the transport coefficients $S,\  \sigma, \ \kappa_e$ and the power factor ($S^2 \sigma$) as functions of the carrier concentration, $n_e$, and compare with the corresponding values of bulk silicon. Fig.~\ref{fig:seebeck} shows the electronic transport coefficients of semiconductor-like Si-I-X systems (blue cells in Table~\ref{table:energyTable}) for the n-doping case at $300$K.
As can be seen from Fig.~\ref{fig:seebeck}, the transport coefficients are significantly altered due to the presence of the interstitials. Both Si-I-Si and -Ge systems with split-interstitials have a smaller $S$ compared to bulk Si, while other interstitials leave $S$ relatively unaltered. Introduction of the interstitials consistently decreases both $\sigma$ and $\kappa_e$. All Si-I-X systems except Si-I-C-hexagonal have similar values of the conductivities; a hexagonal C interstitial yields the lowest $\sigma$ and $\kappa_e$. The combination of $S$ and $\sigma$ yields a small PF for the Si-I-C-hexagonal system as well as Si-I-Si and -Ge systems with the split-interstitials. However, the power factor of Si-I-Si and Si-I-Ge approach the power factor of bulk silicon at high carrier concentrations. 
\begin{figure*}
\includegraphics[width=1.0\linewidth]{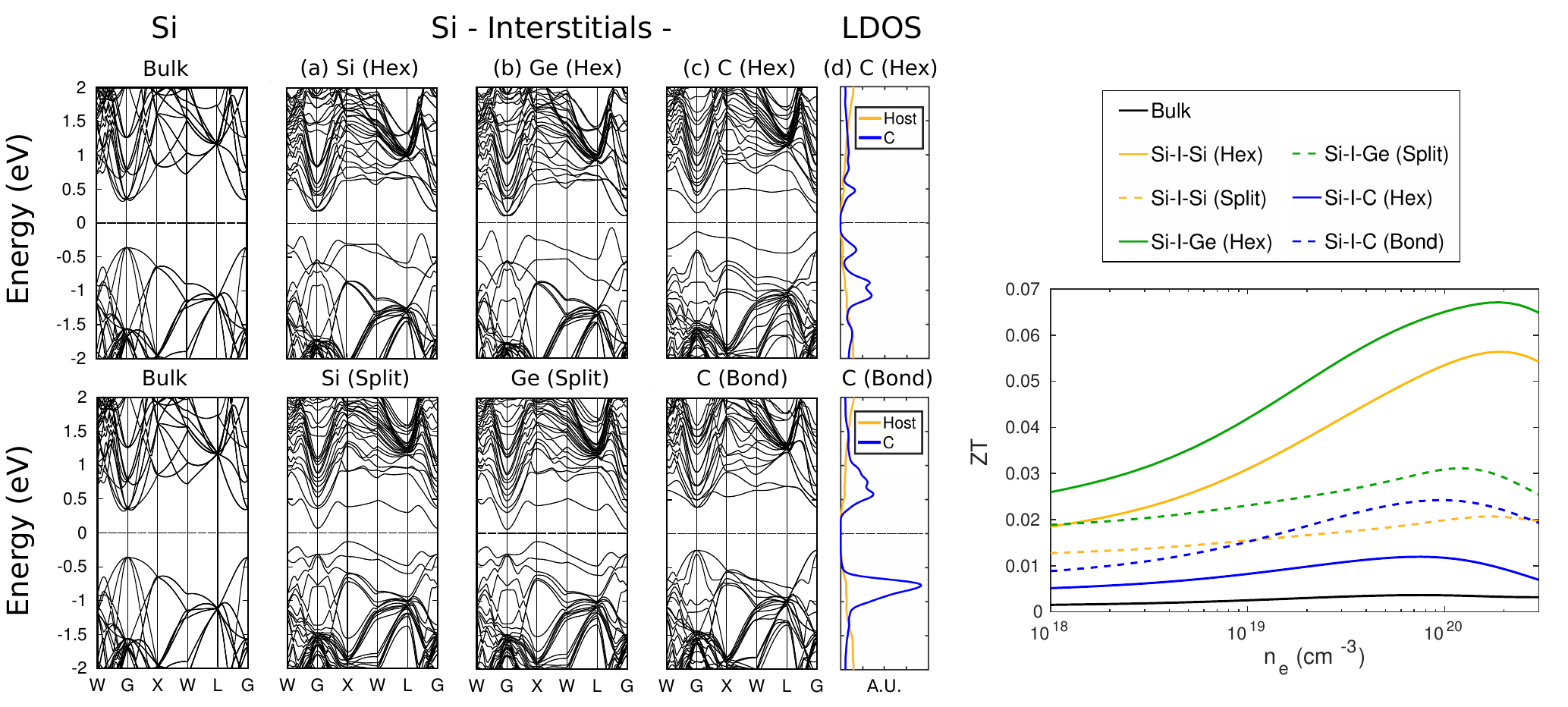}
\caption{Left panel: Band structures of bulk Si and semiconductor-like Si-I-X systems (blue cells in Table~\ref{table:energyTable}). Middle panel: The local density of states (LDOS) of Si-I-C with C interstitial in the bond-centered and the hexagonal sites. The Fermi level is chosen to be at $0$ eV for all cases. Right panel: Figure of merit ($ZT$) of semiconductor-like Si-I-X systems compared to that of bulk Si.}
\label{fig:bands}
\end{figure*}
The electronic transport coefficients were evaluated within the framework of semi-classical Boltzmann transport theory~\cite{ziman1960electrons} as expressed in Eqs. (\ref{eq:L_integral}-\ref{eq:kappa_e}). Therefore, it is essential to examine the factors in the argument of the integral in Eq.~\ref{eq:L_integral}, for a complete understanding of the effect of interstitials on the charge transport coefficients. 

In the following, we discuss the details of the electronic band structures and the density of states (DOS) that determine the transport coefficients. The left panel of Fig.~\ref{fig:bands} shows the band structures of the semiconductor-like Si-I-X systems (blue cells in Table~\ref{table:energyTable}). All interstitial impurities introduce additional levels within the energy bandgap. The levels close to the band edges (either conduction band minimum or valence band maximum) are referred to as ``shallow" levels and those far away from the band edges as ``deep" levels.~\cite{grimmeiss1977deep} The band structures of Si-I-Si (panel (a)) and -Ge (panel (b)) are very similar. Interstitials in hexagonal sites in these two systems introduce two deep levels in the valence zone and few shallow levels in the conduction zone. The same species in the split-interstitial sites create two deep levels in the valence zone with one deep and a few shallow levels in the conduction zone. It is known that the energy gap computed from Kohn-Sham states is systematically underestimated.~\cite{perdew1986density} However, it does not affect the calculation of transport coefficients because we only study transport properties of the highly doped n-type materials, therefore, the integrations in Eq.~(\ref{eq:L_integral}-\ref{eq:kappa_e}) are performed over the energy bands in the conduction zone. The deep levels in the conduction zone decrease transport coefficients due to the $\partial {\it f} / \partial \epsilon$ factor in the integrand as shown in Eq.~(\ref{eq:L_integral}). In general, deep levels lead to poor electron (hole) transport properties because of the low density of available states in the region close to the conduction (valence) zone edge (see Fig.~\ref{fig:bands}). That is the reason the hexagonal interstitials induce better charge transport in the n-type Si-I-Si(-Ge) (Fig.~\ref{fig:seebeck} (solid orange (green) line)) rather than the split-interstitials (Fig.~\ref{fig:seebeck} (orange (green) dashed line)). On the other hand, C-interstitials in hexagonal sites introduce one deep level in both the valence and the conduction zones (panel (c)-top). The deep level in the conduction band leads to poor transport properties as shown in Fig.~\ref{fig:seebeck} (blue solid line). Interestingly, C-interstitials in bond-centered positions do not add any deep levels within the band gap (panel (c)-bottom) and yield better transport properties (Fig.~\ref{fig:seebeck} (blue dashed line)).

In order to understand the origin of the deep and the shallow levels, we compute the local density of electronic states (LDOS) of Si-I-C systems, shown in Fig.~\ref{fig:bands} panel (d). LDOS describes the space-resolved electronic density of states that is computed using postprocessing {\it projwfc.x} code available through QE package. LDOS allows us to quantify the contributions from interstitial and host atoms separately to the full electronic density of states. Figure~\ref{fig:bands} panel (d)-top shows the comparison between the density of states contributed by the C-interstitial (blue) and the host Si atoms (normalized by the number of Si atoms) (orange). The peaks in the LDOS indicate that C-interstitials primarily contribute to the additional deep and the shallow levels, in both hexagonal and bond-center sites. The two peaks (panel (d)-top, blue) in LDOS close to the Fermi level correspond to the deep levels in band structure in Si-I-C-hexagonal systems, while two distinguished peaks $\sim 0.5$ and $\sim 0.7$ eV (panel (d)-bottom, blue) are directly related to the shallow levels in the Si-I-C-bond-centered systems. 

Introduction of the interstitials consistently decreases both $\sigma$ and $\kappa_e$. All Si-I-X systems except Si-I-C-hexagonal have similar conductivities lower than bulk Si, as can be seen from Fig.~\ref{fig:seebeck}; a hexagonal C interstitial yields the lowest $\sigma$ and $\kappa_e$. Lowered conductivities might stem from the additional deep levels which lead to the reduced density of available states in the region close to the conduction zone edge. The combination of $S$ and $\sigma$ yields a small PF for the Si-I-C-hexagonal system as well as Si-I-Si and -Ge systems with the split-interstitials. However, the power factor of the Si-I-Si(and -Ge)-hexagonal systems approach the power factor of bulk silicon at high carrier concentrations. We obtained the thermoelectric figure of merit $ZT$ of the Si-I-X systems at $300$K (Figure~\ref{fig:bands} right panel) by combining the electronic transport coefficients with our MD $\kappa_{\text{ph}}$ results, shown in Fig.~\ref{fig:kappa}.  Since $\kappa_{\text{ph}}$ is strongly suppressed in the presence of the interstitials, we see a significant improvement of $ZT$ for all Si-I-X systems studied in this work compared to bulk Si. The highest $ZT$ are obtained for Si-I-Si and -Ge-hexagonal systems due to a combination of low $\kappa_{\text{ph}}$'s and high PFs. 1.56\% of Ge interstitials in the hexagonal sites are found to improve $ZT$ of bulk Si by a factor of 17. The lowest $ZT$ value is found to be for Si-I-C systems with interstitials in the hexagonal sites. Even for this case the figure of merit is 3 times larger than the reference bulk value \cite{mangold2016optimal}. 

\begin{figure}
\includegraphics[width=1.0\linewidth]{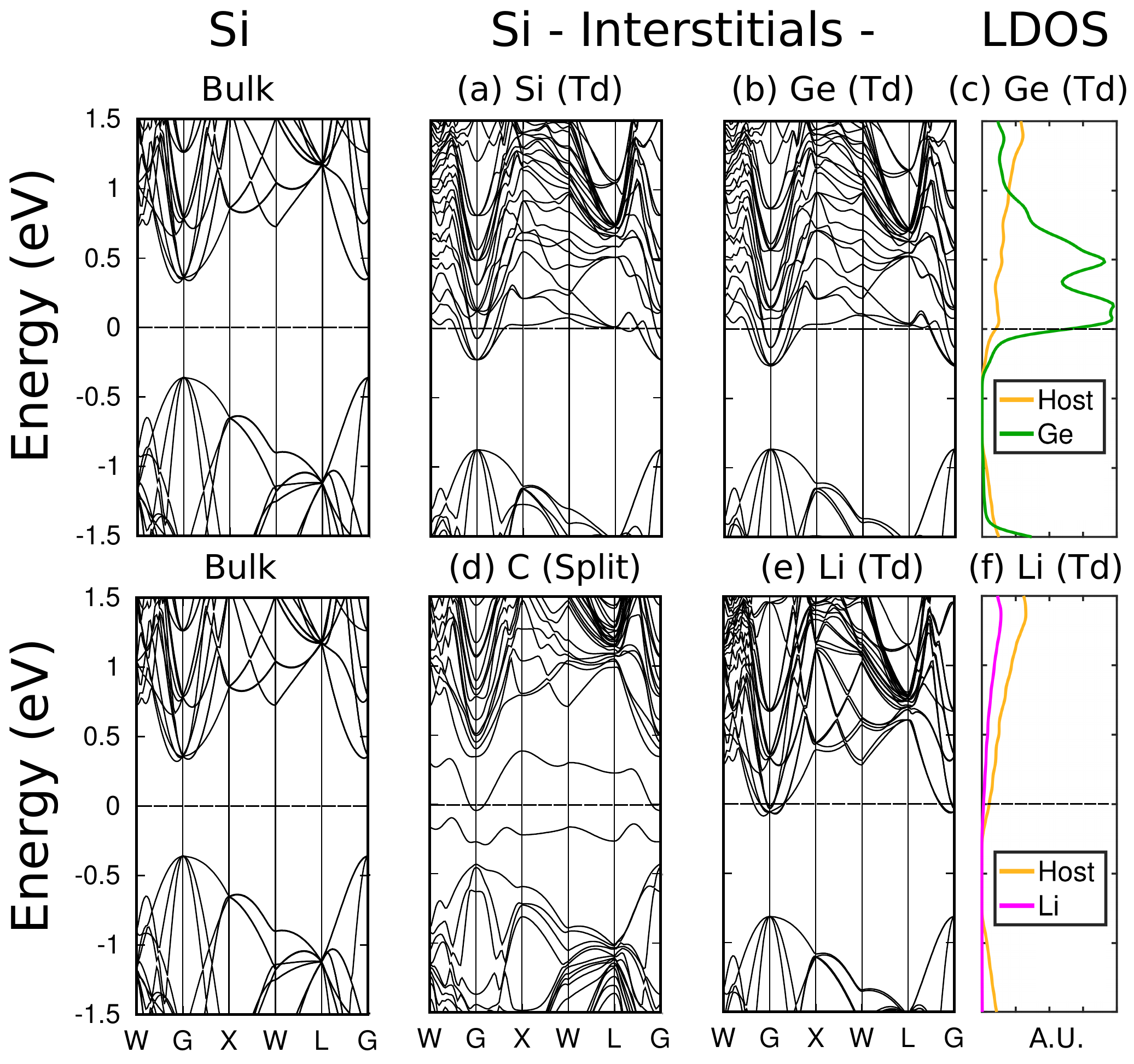}
\caption{ Band structures of bulk Si and metal-like Si-I-X systems (white cells in Table~\ref{table:energyTable}) The plots in the rightmost panels represent the local density of states of (c) Si-I-Ge-tetrahedral and (f) Si-I-Li-tetrahedral systems. The Fermi level is chosen to be at $0$ eV for all cases.}
\label{fig:bs_metal}
\end{figure}
As we discussed in the preceding text, interstitials introduce additional energy levels in the band structure. Both the deep and the shallow levels lead to smaller energy gaps of the Si-I-X systems. We now turn our attention to the systems marked with white cells in Table~\ref{table:energyTable}. The interstitials not only shrink the energy bandgaps of these systems but greatly alter the electronic properties rendering them metal-like. In Fig.~\ref{fig:bs_metal} we present the band structures of the metal-like Si-I-X systems (white cells in Table~\ref{table:energyTable}): Si-I-Si, -Ge and -Li in the tetrahedral sites and -C in the split-interstitial sites. (The interstitial configurations are as depicted in Fig.~\ref{fig:interstitialsType}.) In the Si-I-Si(-Ge or -Li)-tetrahedral systems (panel (a), (b) and (e)) the energy gap is minimally altered but the Fermi level is shifted so that it intersects the conduction bands---such behavior resembles the band structures of highly doped n-type semiconductors. The band structures of Si-I-Si and Si-I-Ge in tetrahedral positions (panel (a) and (b)) are rather similar. The interstitials not only shift the Fermi level but introduce additional shallow energy levels in the conduction zone. In comparison, Li interstitials do not create any additional levels near the conduction band edge (panel (e)). However, the Fermi level is shifted to the conduction zone indicating metal-like behavior similar to Si-I-Si and -Ge in tetrahedral sites. We found no significant difference in the band structures between Si-I-Li-tetrahedral and -hexagonal systems. In order to avoid redundancy, we refrain from displaying the Si-I-Li-hexagonal band structure in this article. The Si-I-C-split-interstitial system exhibits a combined effect of smaller energy gap due to additional deep energy levels (similar to Si-I-C-hexagonal system shown in Fig.~\ref{fig:bands}) and shifted Fermi level leading to metal-like characteristics. 

We present the LDOS of the metal-like Si-I-X systems in the rightmost panel of Fig.~\ref{fig:bs_metal} to discuss the origin of the additional bands. Comparison between the LDOS of Si-I-Ge and Si-I-Li tetrahedral systems (panel (c) and (f)) shows that Ge interstitials (green) are responsible for creating shallow energy levels in $\sim -0.2 -1$ eV range. In contrast, Li interstitials (magenta) do not create distinct peaks either in conduction or valence zones. This can be explained by the fact that Li has fewer electrons than Ge and Si atoms. For the same reason, Li interstitials do not affect the band structure significantly. The orange curves in Fig.~\ref{fig:bs_metal} represent contribution of host atoms to the DOS, scaled with respect to the total number of Si atoms. Additionally, we analyzed the projected density of states in Si-I-Ge tetrahedral system, and found that the additional conduction bands created are mostly formed from the {\it p}-electrons of the guest atom. 
These results indicate that interstitials offer a viable strategy to design n-type semiconductors without explicit dopants, e,g, substitutionally doped bulk silicon with pentavalent impurities such as phosphorus.

\begin{figure}
\includegraphics[width=1.0\linewidth]{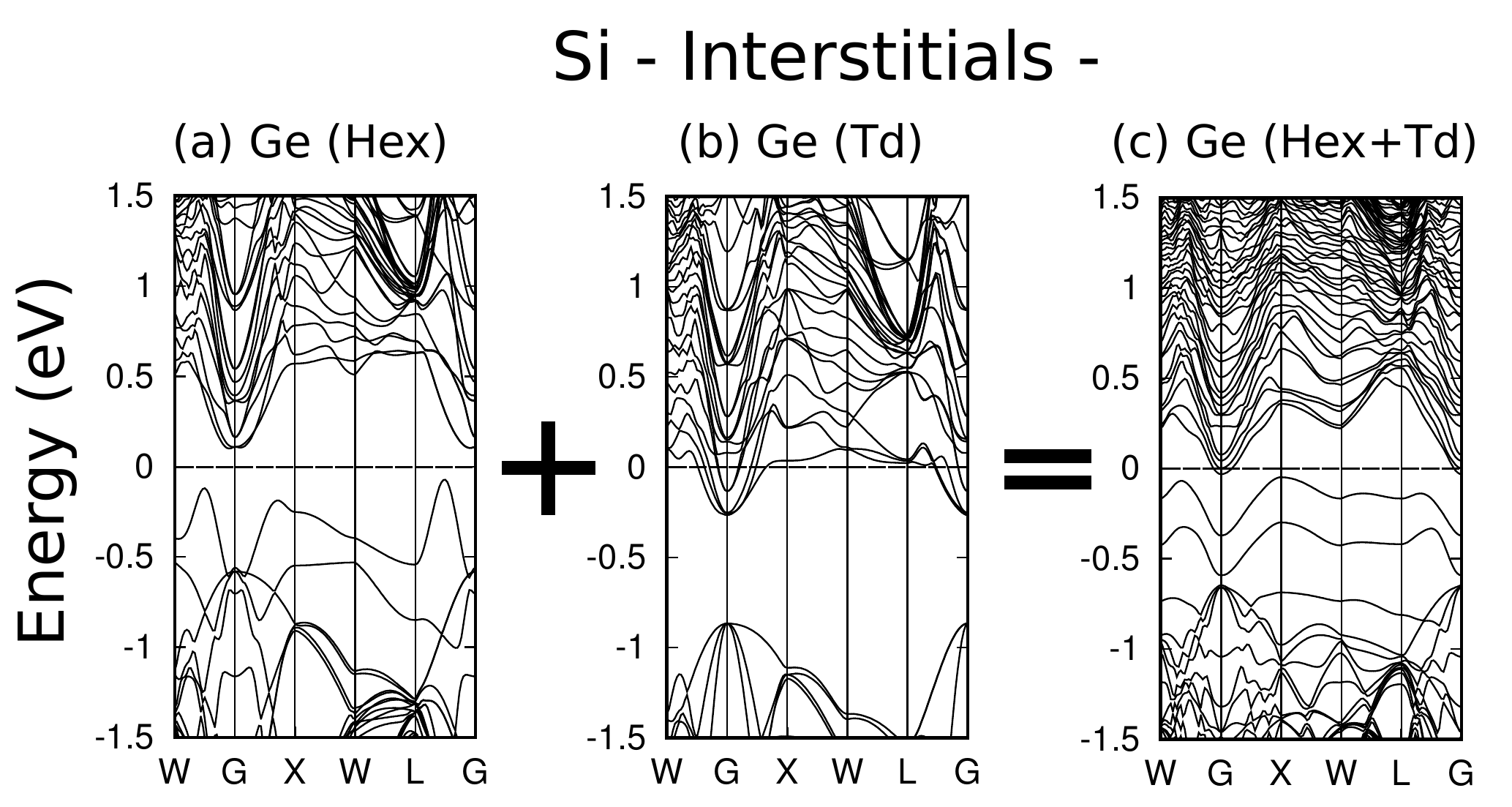}
\caption{  Band structures of Si-I-Ge systems with interstitials located in (a) hexagonal, (b) tetrahedral and (c) both hexagonal and tetrahedral sites. The Fermi level is chosen to be at $0$ eV for all cases.}
\label{fig:bs_ge_comb}
\end{figure}
We acknowledge that in experimental samples the interstitial atoms are likely to reside in different symmetry positions (e.g., hexagonal, tetrahedral) within the crystal due to the variability of the neighbor bonding environment and temperature fluctuations. In order to simulate an experimentally realizable system, we turn to investigate the electronic properties of a Si-I-X system containing interstitial atoms in different symmetry sites. The system consists of 128 Si host atoms and 2 Ge interstitials, one in the hexagonal and the other in the tetrahedral symmetry position. Thus we maintain a 1.56\% concentration of interstitials within bulk, consistent with the cases discussed thus far. In Fig.~\ref{fig:bs_ge_comb} we present the band structures of Si-I-Ge (a) -hexagonal, (b) -tetrahedral and (c) -combination of hexagonal and tetrahedral configurations. As can be noted, the combined system (panel (c)) preserves features from both individual cases (panel (a) and (b)). Hexagonal Ge interstitial introduce two additional levels in the energy gap closer to the valence band edge and additional shallow levels in the conduction zone (panel (a) and (c)), while tetrahedral Ge shifts the Fermi level to conduction zone and introduces additional shallow energy levels in the conduction zone (panel (b) and (c)). To calculate the figure of merit of the Si-I-Ge system with interstitials placed in both hexagonal and tetrahedral positions (panel (c)), we calculate transport coefficients at the Fermi level. Combining results obtained from the electronic transport calculations with our $\kappa_{ph}$ results, we find the figure of merit to be $0.01$.  

\subsection{Structure-$ZT$ map}
\begin{figure}
\includegraphics[width=1.0\linewidth]{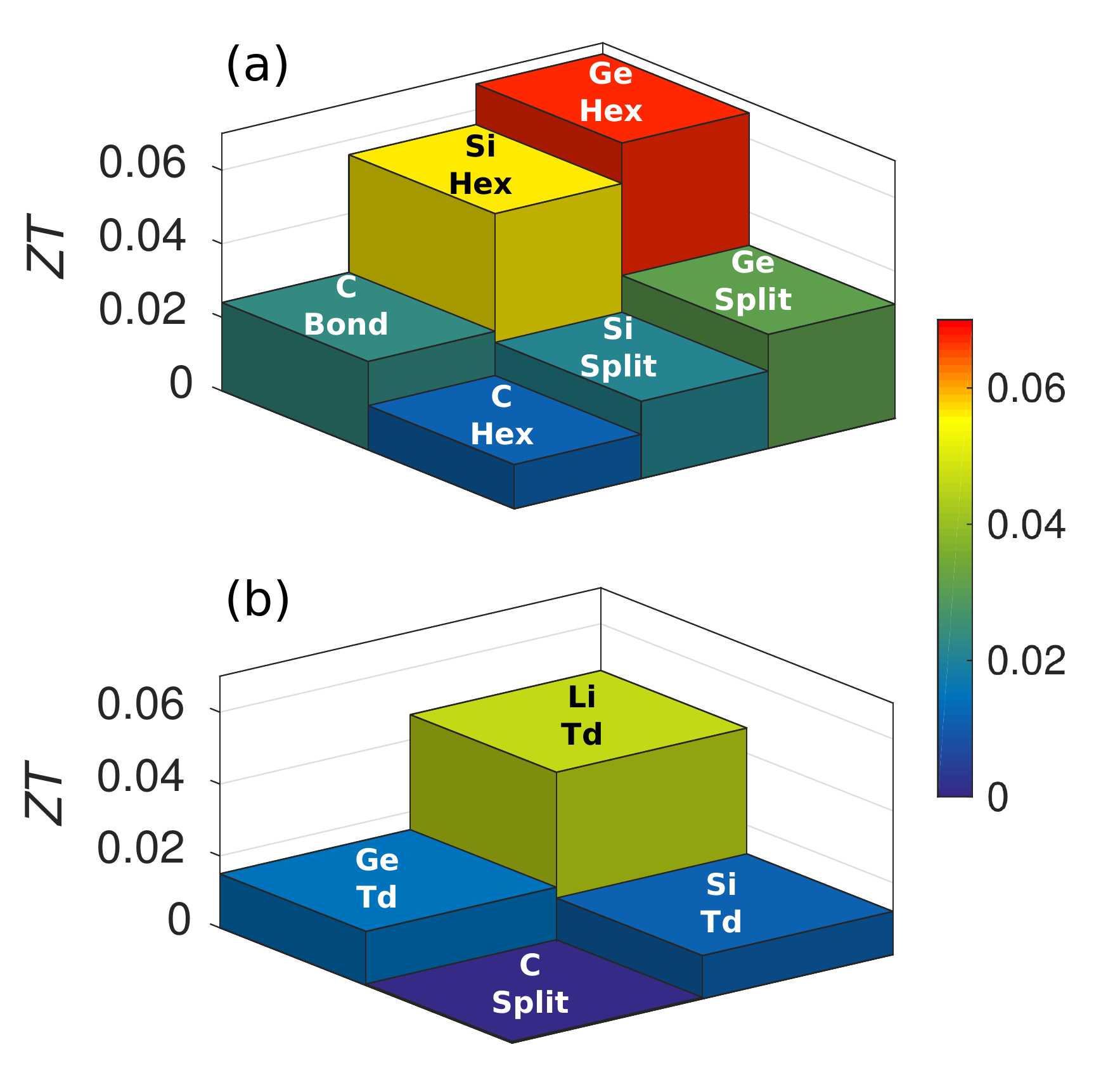}
\caption{Variability of the thermoelectric figure of merit $ZT$ of bulk Si with interstitials, as a function of the guest atom type and its symmetry position in the lattice.}
\label{fig:heatmap}
\end{figure}
We present a comprehensive illustration of the variability of the thermoelectric figure of merit $ZT$ of Si-I-X systems in Fig.~\ref{fig:heatmap}, by combining the electronic transport coefficients with our MD results of $\kappa_{\text{ph}}$. The $x$, $y$ and $z$ axes represent interstitial site type in bulk Si, guest species type and $ZT$, respectively. The top panel (a) displays $ZT$ values of n-doped bulk Si systems with interstitials in different symmetry sites. The carrier concentration is chosen such that the $ZT$ values is optimal. All the systems exhibit higher $ZT$ than bulk Si, which peaks at $\sim$0.004, computed with a similar method~\cite{mangold2016optimal}. The highest $ZT$ value was found for Si-I-Ge-hexagonal systems with  $ZT$ approaching $\sim 0.067$, which is 17 times larger than the reference bulk value. The lowest $ZT$ value $\sim 0.012$ was obtained for Si-I-C system with interstitials in the hexagonal sites. Even for this case the figure of merit is 3 times larger than the reference bulk value. The bottom panel (b) shows $ZT$ of Si-I-X systems with metal-like character. The figure of merit was obtained from the transport coefficients calculated at the Fermi level. The most improved $ZT$ for metal-like systems was found for Si-I-Li-tetrahedral system with $ZT$ value of $\sim 0.047$. The system with C interstitials in split positions yields the lowest $ZT$ value of $\sim 0.0003$, which is 10 times lower than the bulk value.

\section{Summary}

In summary, we performed a systematic study of the effect of naturally occurring randomly distributed interstitial defects (Si, Ge, C and Li) on the charge and heat transport properties of bulk silicon, using first principles DFT and atomistic lattice dynamics and MD techniques, respectively. Our atomistic modeling, using empirical potential, reveals that interstitial defects do not significantly alter the phonon density of states. However, the group velocities, and therefore the propagation of phonons, are strongly suppressed ($\sim$1 order of magnitude). Previous studies reported an additional low frequency peak in the density of states. However, the calculations only incorporated isolated interstitials interacting with simplified neighbor forces, therefore, the results cannot be compared. The reduction of group velocities follows the trend: Ge $<$ Si $<$ C, leading to a decrease in the TCs following the same order: Si-I-Ge $<$ -Si $<$ -C, with Ge having the most and C the least impact in lowering the TC, respectively. Li interstitials are diffusive in character leading to significant disorder in the lattice. This is likely to be the reason for Li interstitials to lower bulk Si TC the most. Randomly dispersed 1.56\% Li, Ge, Si and C interstitials decrease the TC of bulk Si by $\sim 34, \;30, \;20$ and $9$ times, respectively. This suggests that interstitials offer a viable approach to achieve controlled phonon scattering, due to the tunablity of mass and size of the guest atom, and thereby achieve phonon-glass paradigm to engineer next-generation thermoelectric materials. 

In parallel, we investigated the electronic transport properties of bulk Si with 1.56\% Ge, C, Si and Li interstitials in hexagonal, tetrahedral, split-interstitial and bond-centered sites. We demonstrate that Si and Ge in the hexagonal sites minimally impact the charge transport properties of bulk silicon. This combined with the decreased values of $\kappa_{ph}$ leads to 14 and 17 times improved thermoelectric figure of merits without introduction of explicit dopants, respectively. Our results illustrate that electronic transport in defected systems strongly depends on the symmetry positions of defects in the lattice. For example, the bonding environment of split Ge interstitials is marked with the smallest first NN distance 2.3 \AA (obtained from DFT relaxation). The LDOS reveal that the interstitials introduce two deep levels in the valence zone and one deep and few shallow levels in the conduction zone. The deep level (conduction) leads to reduced charge transport properties compared to pristine bulk Si. In comparison, Ge in a hexagonal site, bonded with the first NNs at a distance 2.43 \AA, introduces two deep levels in the valence zone and only shallow levels in the conduction zone. The shallow levels result in a better charge transport. A tetrahedral Ge interstitial, on the other hand, pushes host NN silicon atoms father away to a distance 2.5\AA, yielding a shifted Fermi level resembling a highly doped $n$-type semiconductor. When the Ge interstitial is loosely bound to its first NN, it tends to create additional shallow energy levels. A tightly bound split interstitial leads to deep levels in the band structure that diminishes electron transport. This qualitative understanding, however, may not be generalized to all defects.

We acknowledge that it may not be possible to control the symmetry locations of the interstitials in an experimental sample. We compute the properties of an experimentally realizable silicon system, comprised of both a hexagonal and a tetrahedral Ge interstitial defect. We illustrate that the defects in this system display an additive behavior: the Fermi level is shifted (characteristic of a tetrahedral Ge) as well as additional shallow levels are introduced in the conduction zone (due to hexagonal Ge). Our modeling data furnish indirect measures to estimate the presence of interstitial defects in a sample. 
For example, electronic DOS can be measured with STM techniques.~\cite{yost2016coexistence} The observed band gap and the presence of deep/shallow levels in the STM data can be compared with our theoretical results, to estimate the defects present in the system, and to accurately predict performance of Si-based materials for various technological applications. Our research establishes a direct relationship between the variability of structures dictated by fabrication processes and heat and charge transport properties. We envision that the processing-structure-transport (heat and charge) property map will enable further developments of silicon-based materials with predictable, robust and optimal performance. 

\section{Acknowledgements}
We thank Ty Sterling for a critical reading of the manuscript. This work is funded by the DARPA (DSO) MATRIX program. This work used the Extreme Science and Engineering Discovery Environment (XSEDE), which is supported by National Science Foundation grant number ACI-1548562. We acknowledge the computing resources provided the RMACC Summit supercomputer, which is supported by the National Science Foundation (awards ACI-1532235 and ACI-1532236), the University of Colorado Boulder, and Colorado State University. The Summit supercomputer is a joint effort of the University of Colorado Boulder and Colorado State University.

\bibliography{siLiterature}

\end{document}